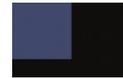

*Article*

# Waveguided Approach for Difference Frequency Generation of Broadly-Tunable Continuous-Wave Terahertz Radiation

**Michele De Regis** [1,*], **Luigi Consolino** [1,2], **Saverio Bartalini** [1,2] and **Paolo De Natale** [1,2]

1. INO, Istituto Nazionale di Ottica-CNR, Largo E. Fermi 6, I-50125 Firenze, Italy; luigi.consolino@ino.it (L.C.); saverio.bartalini@ino.it (S.B.); paolo.denatale@ino.it (P.D.N.)
2. LENS, European Laboratory for Nonlinear Spectroscopy, Via N. Carrara 1, Sesto, I-50019 Fiorentino, Italy
* Correspondence: michele.deregis@ino.it

**Featured Application:** Trace gas sensing, local oscillator for astronomical heterodyne systems, high-accuracy spectroscopy.

**Abstract:** The 1–10 terahertz (THz) spectral window is emerging as a key region for plenty of applications, requiring not yet available continuous-wave room-temperature THz spectrometers with high spectral purity and ultra-broad tunability. In this regard, the spectral features of stabilized telecom sources can actually be transferred to the THz range by difference frequency generation, considering that the width of the accessible THz spectrum generally scales with the area involved in the nonlinear interaction. For this reason, in this paper we extensively discuss the role of Lithium Niobate (LN) channel-waveguides in the experimental accomplishment of a room-temperature continuous wave (CW) spectrometer, with µW-range power levels and a spectral coverage of up to 7.5 THz. To this purpose, and looking for further improvements, a thought characterization of specially-designed LN waveguides is presented, whilst discussing its nonlinear efficiency and its unprecedented capability to handle high optical power ($10^5$ W/cm$^2$), on the basis of a three-wave-mixing theoretical model.

**Keywords:** waveguide; nonlinear optics; terahertz sources

## 1. Introduction

The terahertz (THz) spectral window, generally considered to span from 1 to 10 THz, is nowadays crucial for plenty of normal day-to-day, as well as scientific applications, in areas that include things like homeland security [1], quality inspection of industrial products [2,3], and biomedicine [4]. Among these applications, high-precision molecular spectroscopy and trace gas sensing, requiring powerful and widely tunable continuous wave (CW) sources, play a central role [5]. Indeed, in many cases, the chemical composition of gas mixtures of a very practical interest can be retrieved by observing only a limited part of their spectrum, as long as it contains a sufficient amount of intense transitions serving as a signature for the molecules of interest.

For example, in measuring the concentration of specific chemical species on Earth and astrophysical environments, crucial information can be retrieved about global warming mechanisms [6], star and galaxies formation processes [7], and planetary atmospheres' composition [8].

Because of their strong permanent dipole moment, the rotational transitions of many light molecules, lying in the THz spectral range, are usually comparable in intensity with the strongest ro-vibrational transitions in the mid-IR fingerprint region [9]. Furthermore, the low sensitivity to diffusion losses by micrometric powders and the Doppler linewidths, being one order of magnitude smaller with respect to the mid-IR transitions, make the THz window particularly suitable for the identification of complex line patterns and for in situ measurements [10].

Anyway, despite its undoubted usefulness, the THz interval is still relatively unexploited as a "fingerprint region". This fact is mainly due to the lack of room-temperature and high-power source, able to combine the narrow-linewidth emission required for high-precision spectroscopy (typically in the order of a few kHz) with a spectral coverage and tunability extended to the whole 1–10 THz frequency range.

Actually, emblematic and extreme examples are represented by tunable far-IR lasers (Tu-FIR) which date back to the mid-80s [11–13] and, on the other hand, by THz quantum cascade lasers (THz-QCLs) [14], whose specific features are, in some ways, complementary. Indeed, THz-QCLs are capable of providing THz radiation in the milliwatt range by a cascade of intra-sub-band laser transitions, but their spectral coverage is generally limited to a few percent of the emission frequency [15], which is fixed by the specific device design. Remarkably, at the state-of-the-art level, the maximum frequency achievable with CW THz-QCLs (requiring cryogenic cooling to operate) is limited to 4.75 THz [16,17]. Conversely, an emission of up to 9 THz has been demonstrated with Tu-FIR lasers, taking advantage of a difference frequency generation (DFG) approach [18]. Despite the very low generated power (generally tens of nW, requiring liquid-He cooled detectors), such a wide-tunability approach has produced, in about fifteen years, plenty (several tens) of significant spectroscopic results (e.g., [6,18–23]). However, after a comparable time-span, THz QCLs have provided mostly proof-of-principle spectroscopic results [10,24–26] to date, thus paying the price for insufficient tunability in regard to spectroscopy.

Nonlinear generation of THz radiation by DFG in nonlinear ferroelectric materials, like Lithium Tantalite [27] and Lithium Niobate (LN), provides a suitable alternative with respect to the previous approaches. Indeed, DFG-based setups have been demonstrated to be capable of transferring some of the most interesting features of visible and near-IR sources directly in the far-IR part of the electromagnetic spectrum. For instance, generation of ultrashort THz pulses and comb-like THz emission has been demonstrated by optical rectification (OR) of femtosecond pulses provided by commercial mode-locked lasers [28,29]. On the other hand, the DFG approach can be exploited to transfer the spectral purity and wide mode-hop-free tunability characteristics of the currently available telecom laser systems to the THz range.

Generally, the THz frequency range, achievable with a DFG setup, coincides with the phase-matching (PM) bandwidth of the process, which, in accordance with Heisenberg's uncertainty principle, is limited by the finite interaction area within the nonlinear material [30]. Several different PM geometries have so far been adopted, including tilted-front-pulse generation [31], collinear and non-collinear quasi-PM in periodically-poled LN crystals [29], and Cherenkov generation [32,33]. The intensity-dependent self-beam-defocusing effects in LN [34,35], incompatible with the achievement of an adequate PM, have hampered the possibility to produce broadly-tunable and spectrally pure CW THz radiation by means of bulk nonlinear crystals.

Conversely, by exploiting the confining properties of an MgO-doped LN waveguide, we recently realized an ultra-narrow-linewidth THz spectrometer capable of absolute frequency measurements of molecular transition in the whole 1–7.5 THz range [36]. The current accuracy ($\sim 10^{-9}$), achieved in a direct-absorption configuration, was limited by the Doppler linewidth of the measured absorption profiles. However, improvement of the order of magnitude is possible by implementing more advanced setups [26,37]. To this purpose, higher THz power levels, in the order of a few $\mu$W and not yet available with our source, are desirable.

The goal of this paper is to clarify the centrality of the guided-wave approach in overcoming the main difficulties of CW THz-DFG, and, at the same time, to identify a suitable strategy in order to overcome the existing criticalities, as well as to obtain, if possible, the desired power level. To this purpose, here we provide a comprehensive characterization of the source described in [36], and we discuss its spectral coverage and efficiency on the basis of the three-coupled-wave theory.

## 2. Theory

DFG is a three-wave-mixing process, basically taking advantage of the anharmonic response of a nonlinear medium to an external dual frequency field. Let us consider a laser beam resulting on the superposition of two monochromatic waves at angular frequencies $\omega_1$ and $\omega_2$ (whose difference lies in the THz range $\omega_{THz} = \omega_1 - \omega_2$), where their propagation direction is indicated as $x$. When it passes through a non-centrosymmetric nonlinear medium, it excites a second-order polarization field $P^{(2)}$, which acts as a source of new spectral components and, in particular, of a new field at angular frequency $\omega_{THz}$.

In the most general picture, the three waves $E(\omega_1)$, $E(\omega_2)$, and $E(\omega_{THz})$ are coupled by the three nonlinear wave equations:

$$\nabla^2 E(\vec{r},\omega_i) + \frac{\omega_i^2 n^2(\omega_i)}{c} E(\vec{r},\omega_i) = -\mu_0 \omega_i^2 P^{(DFG)}(\vec{r},\omega_i), \quad (i = 1, 2, THz), \quad (1)$$

where $\mu_0$ is the magnetic permeability of the vacuum, and $n(\omega)$ is the phase velocity of the electric field Fourier component at the angular frequency $\omega$. When the conversion efficiency is sufficiently low and the field amplitude does not vary appreciably within one wavelength (slow-varying-envelope approximation), one can assume the two optical waves to be independent on the THz one. Under these assumptions, the THz field can be described by the nonlinear equation:

$$\frac{dE_{THz}}{dx} = -i \frac{\omega_{THz} d_{eff}}{n_{THz} c} E_1(y,z) E_2^*(y,z) \exp(-i\Delta k x), \quad (2)$$

where $d_{eff}$ is the effective nonlinear coefficient of the medium, and the wavevectors' mismatch $\Delta k$ depends on the refractive indexes for the optical pump fields $n_{pump} \stackrel{\text{def}}{=} n_1 = n_2$, the refractive index for the THz wave is $n_{THz}$, and the angle $\theta$ is formed by the emitted THz wave with respect to the $x$ direction (see Figure 1a): $\Delta k = \frac{\omega_{THz}}{c}(n_{pump} - n_{THz} \cos\theta)$.

Assuming the two pump beams have a Gaussian shape with a $r_0$ radius at the 1/e level, the nonlinear wave equation can be easily integrated over the propagation length within the nonlinear medium $L$, leading to a nonlinear efficiency $\eta$ (defined by the ratio $\frac{P_{THz}}{P_1 P_2}$ between the THz power and the product of the pump beam powers).

$$\eta = \frac{\mu_0 L^2 d_{eff}^2}{c\pi^2 n_{THz} n_{pump}^2 r_0^2} \cdot g(\omega_{THz}, r_0) \cdot \text{sinc}^2\left(\frac{\Delta k L}{2}\right), \quad (3)$$

where:

$$g(\omega_{THz}, r_0) = \omega_{THz}^2 \exp\left(-\frac{\omega_{THz}^2 r_0^2}{4c^2}(n_{THz}^2 - n_{pump}^2)\right). \quad (4)$$

As it can be seen, unless a perfect phase-matching condition ($\Delta k = 0$) is achieved, the THz generation process will take place only within a finite thickness of the nonlinear medium. It is given by the so-called coherence length $L_{coh} = 2\pi/\Delta k$, depending on the optical features of the material and on the wave vectors' geometry. In order to overcome this problem, different techniques have so far been developed. For example, perfect phase-matching in a collinear geometry (i.e., for $\theta = 0$) can be achieved by taking advantage of the velocity mismatch experienced by non-parallel-polarized pump fields in birefringent materials [38]. To maximally exploit the nonlinear properties of certain materials, a parallel arrangement of the three involved polarizations is often required. This is the case for Lithium Niobate (LN), which is one of the best-suited materials for THz DFG because of its strong nonlinearities [39,40], high photorefractive damage threshold [41,42], and small absorption coefficient, in the order of $10^{-4}$ cm$^{-1}$ in the infrared spectral region [43]. In LN-based setups, a viable alternative consists of restoring a quasi-phase matching condition after each coherence length (only few tens of μm) by introducing a periodic modulation of nonlinear susceptibility. The achievement of collinear generation schemes by means of periodically-poled (PPLN) crystals is a fundamental step in order to accomplish optical parametric oscillators [44,45] capable of THz emission with power levels in the order of tens of µW. However, the strong THz absorption of LN [46] and the quasi-PM bandwidth limits the spectral coverage of these sources to few hundreds of GHz around a specific frequency fixed by the PPLN crystal design.

This serious drawback can be completely overcome by realizing the PM in a non-collinear emission scheme within a LN surface waveguide. Indeed, since IR light propagates faster than the THz waves ($n_{pump} < n_{THz}$) [47,48] in such a material, a perfect PM condition can be fulfilled exactly, and a THz wave with a conical wave-front, defined by the relation:

$$\theta_C = \arccos\left(\frac{n_{pump}}{n_{THz}}\right), \tag{5}$$

is coherently generated along the whole crystal length. In LN, the angle $\theta_C$ (called the Cherenkov angle because of the analogy with the light emission by ultra-relativistic particles) is of about 64°, requiring a cladding material with a proper prism shape to be held in optical contact with the emitting surface, in order to avoid total internal reflection of the produced THz radiation. Since the Cherenkov PM results in an infinite coherence length, it entails the experimental need to keep the IR pump beams collimated along the whole crystal length and, at the same time, overlapped on a small interaction area ($\eta \propto L^2/r_0^2$). To this purpose, the most suitable choice is to couple the IR light with the fundamental mode of a tiny channel waveguide. Furthermore, as the waveguide is ion-planted, it lies just beneath the surface of the LN substrate, and the THz radiation can be immediately extracted, minimizing the absorption losses.

Another important advantage rising from the waveguided approach is the reduction of the area of the interacting wave-fronts. This leads to a dramatic increase in the generation bandwidth. Indeed, looking at the frequency dependence of the nonlinear efficiency entirely contained into the function, we can see that while at the low frequencies the efficiency scales as $\omega^2$, the high-frequency behavior is described by a Gaussian profile with full-width-at-half-maximum $\propto 1/r_0$. From a physical point of view, the high-frequency cut-off can be seen as a consequence of the destructive interference between THz waves generated on the crystal surface and the ones generated at a depth of about $\lambda_{THz}/2$. Since LN possesses a refractive index of about 5.2 in the THz range, only the waves generated within a few microns below the surface will experience coherent interference.

In Figure 1b, the function $g(\omega_{THz}, r_0)$ has been plotted for different mode radii. As it can be seen, a broadening of the achievable bandwidth and a high-frequency shift of the maximum available efficiency rapidly occurs once the mode size is reduced. For example, radiation beyond 4.5 THz, representing the current frequency limitation in CW THz QCLs, can hardly be obtained with mode dimensions exceeding 10 µm. Conversely, a beam radius smaller than 5–6 µm should, in principle, be sufficient in order to achieve nonlinear generation along the whole THz bandwidth.

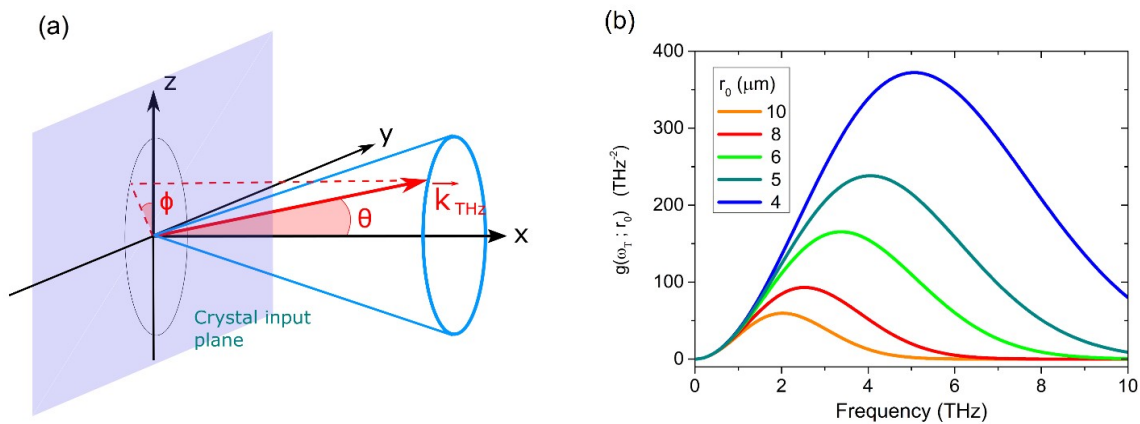

**Figure 1.** (a) Polar coordinates reference system with respect to the waveguide input facet. (b) Accessible THz bandwidth for a continuous wave (CW) waveguided difference frequency generation (DFG) process in MgO-doped Lithium Niobate (LN). Different lines correspond to different guided mode radii between 4 µm and 10 µm.

3. Experimental Results and Discussion

efficiency rapidly occurs once the mode size is reduced. For example, radiation beyond 4.5 THz, representing the current frequency limitation in CW THz QCLs, can hardly be obtained with mode dimensions exceeding 10 μm. Conversely, a beam radius smaller than 5–6 μm should, in principle, be sufficient in order to achieve nonlinear generation along the whole THz bandwidth.

## 3. Experimental Results and Discussion

A schematic of our experimental setup is shown in Figure 2. Two telecom laser diodes (Toptica DL Pro and MOT/DL Pro) emitting wavelengths at about 1.5 μm provided narrow linewidth radiation (tens of kHz on a millisecond timescale), tunable over more than 10 THz by means of a stepping-motor-controlled external cavity. The two monochromatic sources seeded two distinct fiber amplifiers (A1 and A2 in Figure 2). Amplifier A1 (IPG Photonics EAR-10K-C-SF, power up to 11 W) operates between 1540 nm and 1565 nm wavelength, correspondingly to a tunability of about 3 THz. For the other amplifier (A2), two models can be used, operating in two different narrow bands: 1573–1577 nm (IPG Photonics EAR-7K-1575-SF, power up to 7 W) and 1603–1607 nm (IPG Photonics EAR-3K-1605-SF, power up to 3 W), respectively. The two available combinations allow for generation in the 0.97–4.57 THz range and in the 4.54–8.12 THz range, respectively, while the DFG process transfers the spectral properties of the seed telecom lasers to the THz domain.

The two amplified laser beams are overlapped by a non-polarizing beam splitter and focalized on the input facet of a 5% MgO-doped LN crystal plate (HCP Photonics) by means of a 4 mm focal lens. The crystal plate, having a size of 5 × 8 × 10 mm³, contains two planar waveguides (not confining light in the z direction) and nine channel waveguides, ion-planted on the whole of the side which is 1 cm long.

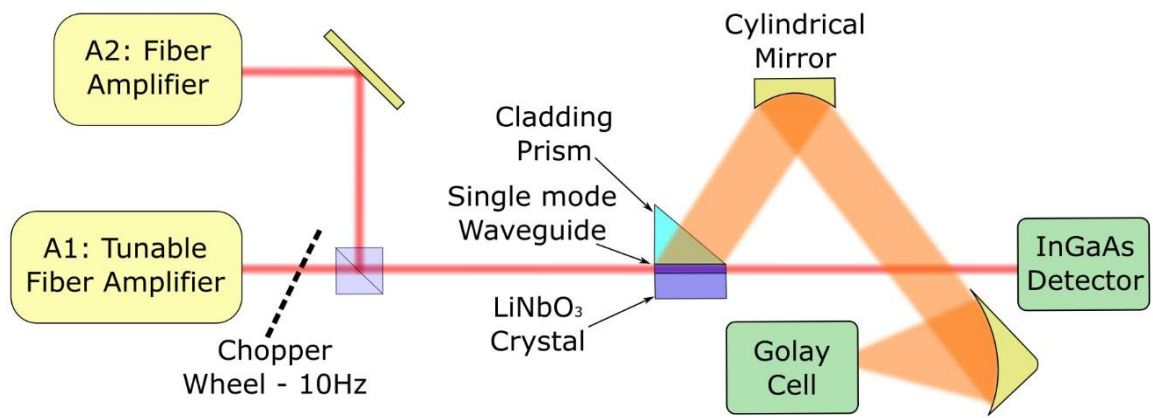

**Figure 2.** Schematic of the experimental setup for waveguided CW THz DFG.

The THz radiation, extracted by means of a 45° cut, high-resistivity silicon (HR-Si) prism, is collimated by a cylindrical mirror and focalized on a room-temperature Golay-type detector (Tydex GC-1P) with a noise-equivalent power of about $10^{-10}\, \frac{W}{\sqrt{Hz}}$ at 10 Hz, and with a calibrated responsivity of 20 kV/W. A 10 Hz mechanical modulation on one of the pump beams, and a fast detector (an InGaAs photodiode with 150 MHz bandwidth—Thorlabs PDA10CF-EC) placed at the waveguide output allowed us to achieve real-time monitoring of the coupled mode powers $P_1$ and $P_2$, and retrieve the nonlinear generation efficiency.

### 3.1. Waveguide Characterization and Power Handling

Knowledge of the geometrical parameters describing the intensity distribution of the guided fields, related to the refractive index gap $\Delta n$ between the core of the nonlinear waveguide and the bulk LN crystal, is crucial for estimation of both the spectral coverage of the CW source (see Equations (3) and (4)) and the maximum IR power available for frequency down-conversion. Indeed, the generation of broadly tunable CW THz radiation, with power levels allowing room-temperature detection, implies the IR pump fields' intensity in the order of $I \simeq 10^7$ W/cm². In these conditions, third-order nonlinear susceptibility and extrinsic defects in the material commonly trigger thermo-optic (TOE) and photo-refractive (PRE) effects, inducing self-defocusing of the pump beams and hampering the achievement of DFG phase-matching [35]. Since the self-defocusing is caused by an implies the IR pump fields' intensity in the order of $I \simeq 10^7$ W/cm$^2$. In these conditions, third-order nonlinear susceptibility and extrinsic defects in the material commonly trigger thermo-optic (TOE) and photo-refractive (PRE) effects, inducing self-defocusing of the pump beams and hampering the achievement of DFG phase-matching [35]. Since the self-defocusing is caused by an intensity-dependent negative modulation of the refractive index ($\Delta n(I) < 0$) between the center and peripheral rays of the Gaussian IR mode, the waveguide refractive index gap has to be larger than $|\Delta n(I)|$.

In order to measure the size of the guided mode, the crystal plate containing the surface waveguides, held by a metallic support, is mounted on a 3 axis platform (Thorlabs MAX313D) allowing a 1 µm accuracy in the positioning of the waveguide input facet with respect to the incoming beam axis. Referring to Figures 1 and 3, we will indicate the propagation direction as the $x$ axis, coinciding with the side of the crystal that is 1 cm long. The plane containing the input facet will be indicated as $yz$, being the vertical $z$ axis parallel to the extraordinary crystallographic axis. The crystal is mechanically held between a copper plate (ensuring optimal heat dissipation) on one side, and the HR-Si prism on the side containing the waveguides. When the pump fields are coupled to the waveguide, the transmitted intensity can be measured as a function of the displacement of the center of the guiding core with respect to the optical axis. Hence, the resulting spatial distribution in the $yz$ plane will be in the form of a convolution that is integral between the intensity profile of the field $I_f$ and of the guided mode $I_g$:

$$T(y,z) = \int I_f(\eta,\varepsilon) I_g(\eta-y, \varepsilon-z) \, d\eta \, d\varepsilon \tag{6}$$

Generally, the guided mode intensity profile $I_g$ can undergo an exponential or a Gaussian decay at the edge of the guiding core, depending on the fabrication process [49]. In the following, we will assume for it a Gaussian profile, since the convolution integral obtained assuming an exponential decay did not reproduce our experimental data. On the contrary, as it will be shown, they are correctly reproduced by assuming a Gaussian shape for both the field intensity $I_f(y,z) = I_{f0} \exp\left(-\frac{2(y^2+z^2)}{w_0^2}\right)$ and the guided mode $I_g(y,z) = I_{g0} \exp\left(-2\left(\frac{y^2}{r_{0y}^2} + \frac{z^2}{r_{0z}^2}\right)\right)$. In so doing, even the transmitted power profile will be described by a Gaussian with radii at the 1/e level:

$$\sigma_j = \frac{1}{2}\sqrt{w_0^2 + r_{0j}^2} \quad (j = y, z).$$

The measurement of the waist of the input beam $w_0$ is non-trivial. Indeed, since an optimal mode coupling requires $w_0$ to be of the same order of magnitude of the mode size (namely, a few microns), it cannot be measured by means of the standard knife-edge method [50]. Our approach consists in acquiring different transmission profiles for different positions of the waveguide facet with respect to the focal plane. According with the Gaussian beam propagation, the acquired profiles will be broadened, increasing the relative displacement $\Delta x$ between the two planes. The transmitted intensity radius can be expressed as:

$$\sigma^2 = a + b \cdot (\Delta x)^2$$

where the guided mode radius is related to the linear parameters $a$ and $b$ by:

$$r_0 = \sqrt{\frac{2}{b}}\sqrt{ab - \left(\frac{\lambda}{2\pi}\right)^2}. \tag{7}$$

Such a broadening effect can be clearly observed in the different experimental traces of Figure 3b,c, representing the transmitted profiles in the $y$ and $z$ directions, respectively. It must be noted that, while the Gaussian shapes obtained in the $z$ direction are symmetric for every value of $\Delta x$, the LN-Silicon interface implies a visible decrease in guiding capability in the $y > 0$ half-space.

Furthermore, since only the volume corresponding to $y < 0$ is involved in the generation process, the data corresponding to positive values of $y$ have not been fitted.

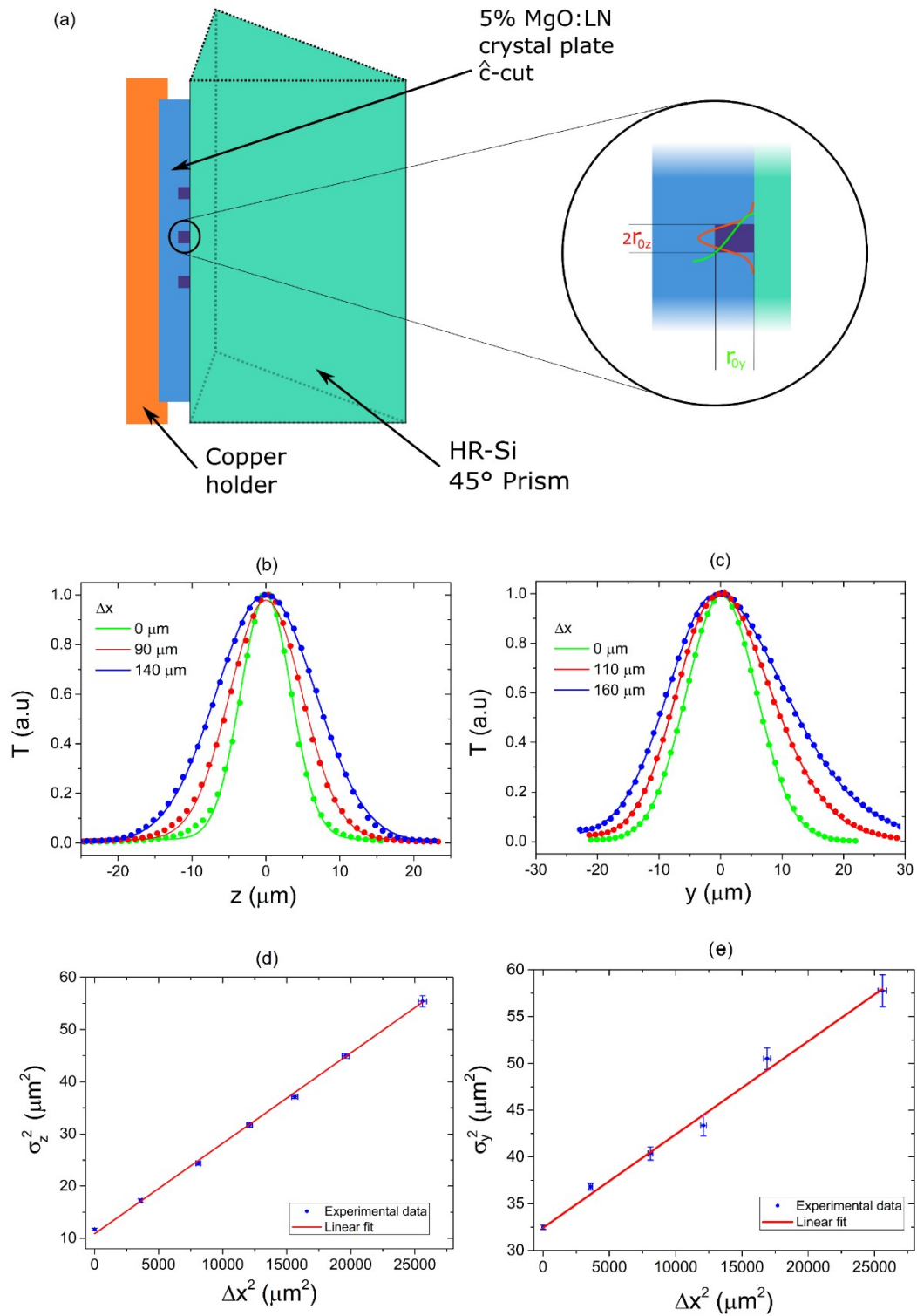

**Figure 3.** (**a**) Representation of the crystal plate containing the channel waveguides (**b**,**c**) Transmission profiles in the $z$ and $y$ directions, respectively, measured for different waist displacement $\Delta x$. (**d**,**e**) Mode Radii at the 1/e level as a function of $\Delta x$.

The measurement of the waist of the input beam $w_0$ is non-trivial. Indeed, since an optimal mode coupling requires $w_0$ to be of the same order of magnitude of the mode size (namely, a few microns), it cannot be measured by means of the standard knife-edge method [50]. Our approach consists in acquiring different transmission profiles for different positions of the waveguide facet with respect to the focal plane. According with the Gaussian beam propagation, the acquired profiles

In Figure 3d,e, the $\sigma^2$-parameter has been plotted as a function of the squared waist displacement $\Delta x^2$, while the measured waveguide parameters have been summarized in Table 1. The radii $r_{0y}$ and $r_{0z}$ of the guided mode have been obtained by substituting the fitting parameters into Equation (7), and the corresponding uncertainties have been calculated by means of the standard propagation formula. Knowledge of the intensity distribution $I_g$ allows us to estimate the refractive index gap $\Delta n$ between the guiding material and the substrate. It has been calculated on the basis of the measured geometrical parameters by applying the inverse WKB method [51]. In the weak guidance approximation ($\Delta n \ll n_{pump}$), the n-gap can be obtained by evaluating the expression:

$$\Delta n(y,z) = -\frac{\nabla^2 \sqrt{I_g(y,z)}}{2 n_{pump} k_0^2 \sqrt{I_g(y,z)}} \qquad (8)$$

In the guiding core ($y = z = 0$). In the previous expression, $k_0$ represents the free-space IR wavevector, while for $n_{pump}$ the value provided by the Sellmeier semi-empirical formula has been used [47].

Table 1. Measured waveguide parameters.

| $r_{0y}$ | $r_{0z}$ | $n_{pump}$ | $\Delta n$ |
|---|---|---|---|
| $(8.3 \pm 0.2)$ μm | $(2.95 \pm 0.27)$ μm | 2.13 | $(4 \pm 1) \times 10^{-3}$ |

The obtained result of $\Delta n = 0.004$ has been compared with the contribution, opposite in sign, due to the thermo-optic and photo-refractive effects. These quantities have been estimated from the relations [52]:

$$\Delta n_{TOE} = -\frac{dn}{dT} \frac{\alpha r_0^2 I}{k}$$

and:

$$\Delta n_{PRE} = -n_{pump}^{5/2} r_{33} \left(\frac{I}{2c\varepsilon_0}\right)^{1/2} \qquad (9)$$

providing a contribution in the order of $10^{-6}$ and $10^{-4}$, respectively. In the previous equations $\alpha$, $k$ and $r_{33}$ are the absorption coefficient, the thermal conductivity, and the electro-optic coefficient of the LN crystal, respectively, whose values used for calculations are summarized in Table 2. When compared with the previous ones, the measured value $\Delta n$ predicts, for our waveguide, the capability to confine an IR intensity of up to $10^6$ W/cm$^2$ in the guided mode.

Table 2. Characteristic parameters of LN crystal [52].

| $dn/dT$ | $\alpha$ | $k$ | $r_{33}$ |
|---|---|---|---|
| $3 \times 10^{-5}$ K$^{-1}$ | $10^{-4}$ cm$^{-1}$ | 5 Wm$^{-1}$K$^{-1}$ | 30 pm/V |

The experimental curve, printed in blue in Figure 4a, represents the measured values of the coupled vs. incoming IR power. For the sake of clarity, these data have been compared with an analogous measurement performed by confining light in the fundamental mode of a planar waveguide (red trace in Figure 4a). Such a waveguide is characterized by the same geometrical and optical properties of the channel one, except for the radius of the beam in the non-guiding direction, which, in this case, is $b_z = 150$ μm. The measurements here reported show how, in the absence of a guiding structure, the coupling efficiency rapidly decreases as the input power exceeds the value of about 1 W, correspondingly to an optical energy density of $1.3 \times 10^5$ W/cm$^2$. Conversely, using a channel-shaped waveguide, up to 3.6 W of IR power (corresponding to $10^6$ W/cm$^2$ intensity) have been coupled to the waveguide (red trace in Figure 4a). Such a waveguide is characterized by the same geometrical and optical properties of the channel one, except for the radius of the beam in the non-guiding direction $z$, which, in this case, is $r_{0z} = 150$ μm. The measurements here reported show how, in the absence of a guiding structure, the coupling efficiency rapidly decreases as the input power exceeds the value of about 1 W, correspondingly to an optical energy density of $1.3 \cdot 10^5$ W/cm$^2$. Conversely, using a channel-shaped waveguide, up to 3.6 W of IR power (corresponding to $10^7$ W/cm$^2$ intensity) have been coupled to the guided mode, with a constant coupling efficiency of about 57.3%. This result demonstrates the channel waveguide to be a fundamental tool in high-power frequency down-conversion experiments.

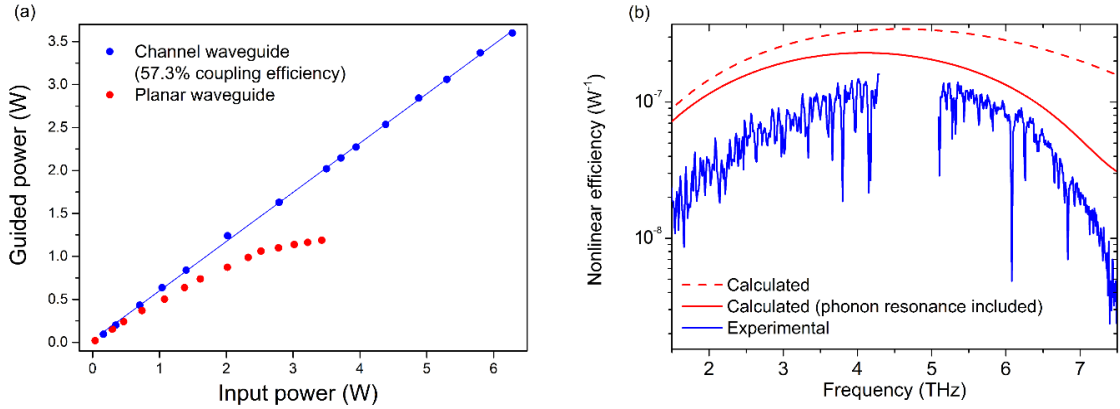

**Figure 4.** (a) Infrared power coupled to the guided mode as a function of the incoming power for a channel waveguide (blue) and a planar waveguide (red). (b) The red curves represent the calculated nonlinear efficiency, $\eta$, with (solid) and without (dashed) the contribution of the transverse optical phonon resonance at 7.5 THz. The blue curve is an example of the experimental nonlinear efficiency obtained with two different frequency scans. The low- and the high-frequency parts of the blue trace have been obtained by scanning the A1 amplifier from 1541 to 1563.5 nm and fixing the A2 wavelength at 1575 nm and 1605 nm, respectively.

### 3.2. THz Generation Efficiency and Spectral Coverage

The generation bandwidth achieved by our CW THz source has been investigated by tuning the wavelength (i.e., the frequency $\omega_1$) of one of the laser telecom diodes along the 1540–1635 nm operational bandwidth of the A1 amplifier, while the frequency $\omega_2$ of the other laser is fixed within the A2 bandwidth. As previously discussed, two different combinations are possible, allowing us to span two different portions of the THz spectral window. Figure 4b shows two examples of these frequency scans, one for each amplifier's configuration, demonstrating a spectral coverage extended from 1 to 7.5 THz. The first half of the curve, spanning the 1 to 4.5 THz spectral window, has been obtained by fixing $\lambda_2 = 2\pi c/\omega_2$ at 1575 nm and adjusting $\lambda_1$ from 1541 nm to 1563.5 nm. In this range, a maximum THz power of 0.52 μW has been measured at 4.1 THz. This value, corresponding to 3.6 W of IR power confined in the waveguide (distributed in $P_1 = 1.8$ W and $P_2 = 1.8$ W), leads to a nonlinear efficiency of $\eta = 1.5 \times 10^{-7}$ W$^{-1}$. The second part of the blue trace in Figure 4b, from 5 to 7.5 THz, has been achieved with the same $\lambda_1$ scan and setting $\lambda_2 = 1605$ nm. Here, a maximum THz power of 0.2 μW (with $P_1 = 1.8$ W and $P_2 = 0.75$ W) has been measured at 5 THz, correspondingly, to the same nonlinear efficiency.

The operation of a room-temperature setup, spanning the entire 1–7.5 THz frequency range (i.e., almost 3 octaves and well beyond the 4.5 THz current limit of THz-QCLs) with a fraction of μW generated power can represent a groundbreaking achievement for a number of applications. Nevertheless, some hints for further improvements can be found by comparing the experimental generation curve with the expected one.

To this purpose, the nonlinear efficiency $\eta$ has been calculated by numerical integration of Equation (3) over the emission angles. To this regard, given the asymmetric profile of the guided mode ($r_{0y} > r_{0z}$), Equation (4) has been generalized to:

$$g(\omega_{THz} . r_{0y}, r_{0z}) = \omega_{THz}^2 \ \exp(-\frac{\omega_{THz}^2 \ n_{THz}^2}{4c^2}(r_{0y}^2 \sin^2 \varphi \sin^2 \theta + r_{0z}^2 \sin^2 \theta \cos^2 \varphi)). \qquad (10)$$

Since for $L = 1$ cm, the term $sinc^2(\Delta k \ L/2)$ in Equation (3) is sharply peaked around $\Delta k = 0$, $\theta$ has simply been evaluated at the Cherenkov angle, while the integration on $\varphi$ has been performed within the range allowed for transmission across the Si-air interface. The angular dependence has also been included to take into account the Fresnel losses at both the LN-Si and Si-air interfaces and in a multiplicative term $\sin^2 \varphi$, in analogy with a linear dipole emission. In order to correctly reproduce the spectral profile of the measured efficiency, frequency dependence of $f_{THz}$ and of the HR-Si have been considered. In particular, the dispersion curve for 5% MgO:LN has been obtained assuming the Sellmeier equation to be valid in the whole 1–8 THz spectral window, while the absorption spectrum of HR-Si has been retrieved from the THz-bridge spectral database [53]. The numerical result of such a calculation is represented by the red dashed line in Figure 4b. Despite the abundance of considered parameters, a significative discrepancy between the predicted values and measured ones can be noticed. In particular, the high frequency cut-off at 7.3 THz, limiting the achieved spectral coverage, is not reproduced by the simulation. In our opinion, such a discrepancy can be ascribed to the THz absorption in LN, which was so far neglected because the generated radiation went through a very small thickness in the surface waveguide. Remarkably, the absorption profile of MgO:LN is dominated by a transverse-optical phonon resonance centered at 7.5 THz, being strongly dependent on crystal temperature and the MgO-doping concentration. In order to include such an effect in our simulation, the absorption coefficient has been calculated by combining the four-resonances phonon-polariton dispersion model developed by Bakker et al. [54], with the measured parameters provided by Unferdorben et al. for 6.1% MgO:LN at room temperature [46].

A significant improvement in the matching between the predicted and observed efficiency is represented by the red solid curve in Figure 4b. This suggests that the presence of the phonon mode in MgO:LN plays a dominant role, limiting the maximum achievable spectral coverage of our source, even for less than 10 µm THz propagation within the nonlinear medium. A residual discrepancy at the higher frequencies can still be noticed. In our opinion, this could be ascribed to the presence of a thin air gap $\Delta$ between the waveguide surface and the silicon prism. Indeed, since the Cherenkov angle is not sufficient to allow for THz transmission at the LN-air interface, in this case the amplitude of the extracted THz wave would be attenuated by an exponential factor $e^{-\beta \Delta}$, where the cut-off parameter increases as $\beta \propto n_{THz}^2 \omega_{THz}$ [55]. In our setup, the high IR power levels used in the THz generation process hamper the possibility to use an additional matching material between the LN crystal and the prism. For this reason, we are not comfortable in providing a quantitative value for the cut-off frequency, the $\Delta$ parameter being substantially unknown. It is important to stress that, in principle, even an air gap in the order of a few hundreds of nm could be sufficient to justify both the high-frequency limitation and the overall efficiency of the CW THz source.

The previous analysis suggests considerable room for improvement, both in terms of absolute THz power and of spectral coverage. In particular, the dimensions of the optical guiding structure allow for a PM bandwidth as wide as 1–8 THz, that can be further enhanced by optimizing the optical contact between the waveguide surface and the HR-Si prism. At the same time, the phonon-induced THz absorption can be minimized by inducing a fast thermo-electric heat dissipation of the LN crystal, preserving the room-temperature operation of our source.

## 4. Conclusions

In this work, the central role of nonlinear optical waveguides in the Cherenkov generation of a broadly tunable CW THz radiation has been extensively discussed. In particular, a comprehensive experimental characterization of a MgO:LN waveguide has been carried out, and its capabilities to achieve unprecedented optical power confinement (with an intensity in the order of $10^7$ W/cm$^2$) have been demonstrated. Moreover, the analytical expression for the nonlinear generation efficiency has been derived on the basis of a general three-coupled-waves model. Such a calculation, combined with the guided mode-size experimental measurement, provided a close comparison between the experimental generation curve and the simulated one. In this way, the main technical criticalities and possible improvements of our CW THz source have been pointed out, suggesting a phonon-induced absorption and low extraction efficiency at the LN-Si interface as the main causes of the frequency roll-off and limitations in THz power.

The measurements carried out in this work also suggest that our source, in principle, could emit several microwatts of CW THz power, by optimizing the waveguide heat dissipation (through a simple thermo-electric cooling system) and surface-matching with the cladding substrate. The achievement of these power levels, fostering more sophisticated sub-Doppler spectroscopic setups, would represent a breakthrough for THz science, opening the way to measurements of unprecedented accuracy in the 1–8 THz range.

**Author Contributions:** Conceptualization, P.D.N.; methodology, M.D.R. and L.C.; software, M.D.R., L.C. and S.B.; formal analysis, M.D.R. and L.C.; investigation, M.D.R., L.C. and S.B.; data curation, M.D.R.; original draft preparation, M.D.R.; review and editing, L.C., S.B. and P.D.N.; visualization, M.D.R.; supervision, P.D.N.

**Acknowledgments:** Authors would like to acknowledge support from: European Union FET-Open grant 665158 "Ultrashort Pulse Generation from Terahertz Quantum Cascade Lasers"—ULTRAQCL Project; European ESFRI Roadmap "Extreme Light Infrastructure"—ELI Project; European Commission–H2020 Laserlab-Europe, EC grant agreement number: 654148.

**Conflicts of Interest:** The authors declare no conflict of interest.